\documentclass[article,twocolumn,amsmath,amssymb,showpacs]{revtex4}

\usepackage{graphicx}
\usepackage{dcolumn}
\usepackage{bm}
\usepackage{mathrsfs}
\usepackage{appendix}
\usepackage{bbm}
\usepackage{color}

\def \mB {\mu_{\rm B}}
\def \vz {{\bf z}}
\def \vx {{\bf x}}
\def \vS {{\bf S}}
\def \vh {{\bf h}}

\begin{document}

\title{A Spiral Spin State with Open Boundary Conditions in a Magnetic Field \footnote{Copyright notice: This manuscript has been authored by UT-Battelle, LLC under Contract No. DE-AC05-00OR22725 with the U.S. Department of Energy. The United States Government retains and the publisher, by accepting the article for publication, acknowledges that the United States Government retains a non-exclusive, paid-up, irrevocable, world-wide license to publish or reproduce the published form of this manuscript, or allow others to do so, for United States Government purposes. The Department of Energy will provide public access to these results of federally sponsored research in accordance with the DOE Public Access Plan (http://energy.gov/downloads/doe-public-access-plan).}}

\author{Randy S. Fishman and Satoshi Okamoto}
\affiliation{Materials Science and Technology Division, Oak Ridge National Laboratory, Oak Ridge, Tennessee 37831, USA}

\begin{abstract}

In order to model a spiral spin state in a thin film, we study a classical Heisenberg model with open boundary 
conditions.  With magnetic field applied in the plane of the film, the spin state becomes ferromagnetic above a critical
field that increases with thickness $N$.  For a given $N$, the spiral passes through states with
$n= n_0$ up to 0 complete periods in steps of 1.  These numerical results agree with earlier analytic results
in the continuum limit and help explain the susceptibility jumps observed in thin films.

\end{abstract}


\maketitle

Due to its close connection with multiferroic behavior, spiral spin order \cite{Khomskii06, Cheong07} such as in Fig.1 has been the subject of 
intense investigation.  Recently, spiral order was discovered \cite{Wang00, Togawa12, Wilson13, Chapman14, Meynell14, Porter15, Togawa15} in 
several thin films.  In order to manipulate the magnetic properties of these spiral states, it is essential to understand how they
depend on film thickness $N$ and magnetic field $B$.  A spiral in a thin film can be modeled 
by a linear Heisenberg model with neighboring spins coupled by a ferromagnetic interaction $J$ and by a Dzyaloshinskii-Moriya (DM) interaction $D$
\cite{Dzal59, Moriya60} due to broken inversion symmetry.  

\begin{figure}
\includegraphics[width=8cm]{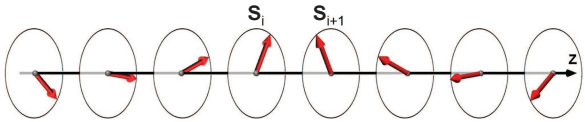}
\caption{(Color) A spiral spin state propagating along $\vz$.}
\end{figure}

In the bulk limit, this simple model has an analytic solution that supports solitonic states \cite{Kishine05}.  
Consequently, the spiral state is often refered to as a ``solitonic" lattice.  For thin films, this model is believed to 
describe materials like Cr$_{1/3}$NbS$_2$ \cite{Togawa12, Chapman14, Togawa15} and MnSi \cite{Wilson13}.
In Cr$_{1/3}$NbS$_2$, Togawa {\it et al.} \cite{Togawa15} directly imaged the ``solitonic lattice" by Lorentz microscopy in a 1 $\mu $m thick sample.  They also
found that the spiral state produces steps in the magnetoresistance versus field, applied perpendicular to the chiral axis or in the plane of the spiral.
In MnSi \cite{Wilson13}, the spiral was detected both in magnetoresistance measurements and in the magnetization versus applied field for 
25 and 30 nm thick samples.  

\begin{figure}
\includegraphics[width=7cm]{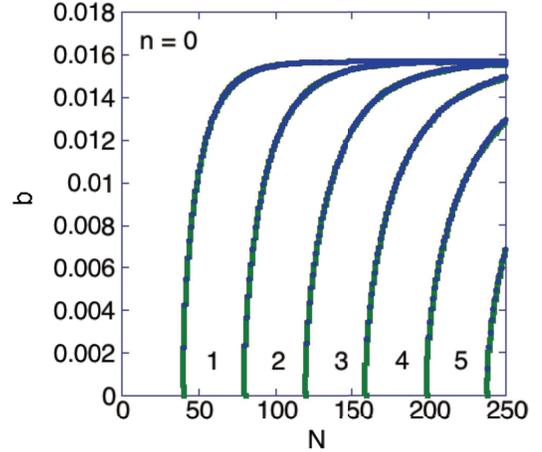}
\caption{(Color) Phase diagram of spiral spin state with $d=0.16$.  Green squares were evaluated by sweeping the magnetic field, blue circles
by sweeping the thickness.}
\end{figure}

For both materials, measurements exhibit discontinuous changes as the number $n$ of periods of the spiral decreases by one with increasing field.  
A high enough field stabilizes a ferromagnetic state with $n=0$ so that all spins point along the field direction.  
Because the chirality of the spiral is maintained against defects and temperature, thin films that
support spiral states have been considered as magnetic storage devices \cite{Braun05}.

To understand the behavior of a spiral sandwiched between two other magnetic materials, 
the model described above has been studied 
\cite{Kishine14} with fixed boundary conditions so that the spins $\vS_i$ on sites $i=1$ and $N$ are fixed along the field direction perpendicular to the chiral axis.  
However, most thin films may be better described using open boundary conditions so that the end spins are free to rotate.  For example, both the 
Cr$_{1/3}$NbS$_2$ and MnSi samples discussed above were grown on magnetically inert Si wafers.

Our numerical solution of this problem qualitatively agrees with an earlier analytic solution \cite{Wilson13} in the continuum limit.  
We predict that thin films will exhibit jumps in the magnetization and peaks in the susceptibility when $n$ 
decreases by one with increasing field.  Remarkably, our results are also quite similar to earlier results that imposed 
fixed boundary conditions \cite{Kishine14}.

The Hamiltonian for this problem is  
\begin{eqnarray}
&&{\cal H} = -J \sum_{i=1}^{N-1} \vS_i\cdot \vS_{i+1} -D \sum_{i=1}^{N-1} \vz \cdot (\vS_i \times \vS_{i+1})\nonumber \\
&&+A \sum_{i=1}^N (\vz \cdot \vS_i)^2 -2\mB B\sum_{i=1}^N \vx \cdot \vS_i,
\label{Ham}
\end{eqnarray}
where the spins $\vS_i $ are treated classically and interact with their neighbors through the ferromagnetic exchange $J$ and the DM interaction
$D$.  The anisotropy $A>0$ keeps the spins in the $xy$ plane.
Notice that the end spins at sites $i=1$ and $N$ are treated differently than spins in the interior.  For example, the
spin at $i=N$ experiences an exchange interaction with the spin at $i=N-1$ but the spin at $N+1$ is missing.
All spins experience the same magnetic field along $\vx $ in the plane of the film.

The dimensionless parameters of this model are $d\equiv D/J$ and $b\equiv 2\mB B /JS$.  Aside from keeping all the spins
$\vS_i = S(\cos \theta_i ,\sin \theta_i,0)$ in the $xy$ plane, the anisotropy $A $ plays no role in the solution to this model.
For a bulk system ($N\rightarrow \infty $), the DM interaction produces a spiral with wavelength $\Lambda =2\pi /\tan^{-1}(d)\approx 2\pi/d $.
When $d>0$, the spiral is right handed;  when $d<0$, it is left handed.

Starting with a uniform spiral, the spin state is allowed to relax in discrete time steps.  
At each step, we obtain updated values for the spins from the condition that $\vS_i$ lies along
the effective field $\vh_i$,
where the energy at site $i$ is given by $-JS\vh_i \cdot \vS_i$.  The dimensionless effective field $\vh_i$ has  
components
\begin{equation}
h_{ix}=\cos \theta_{i+1} + \cos \theta_{i-1} +d \bigl( \sin \theta_{i+1} - \sin \theta_{i-1} \bigr) + b,
\end{equation} 
\begin{equation}
h_{iy}=\sin \theta_{i+1} + \sin \theta_{i-1} -d \bigl( \cos \theta_{i+1} - \cos \theta_{i-1} \bigr).
\end{equation} 
Of course, $h_{iz}=0$.  For the end spins at $i=1$ or $N$, the terms to the left ($i=0$) or the right ($i=N+1$) are missing.

\begin{figure}
\includegraphics[width=8.5cm]{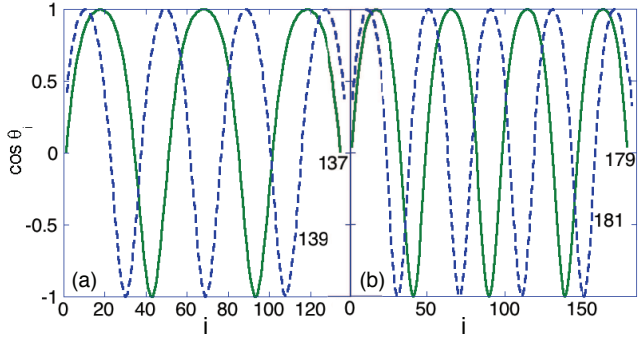}
\caption{(Color) Spin along the field direction as a function of site $i$ for different thicknesses and $b =0.01$.  
In (a), $N=137$ and 139 correspond to $n=2$ and $3$, respectively.  In (b), $N=179$ and 181 correspond
to $n=3$ and 4, respectively.}
\end{figure}

\begin{figure}
\includegraphics[width=8cm]{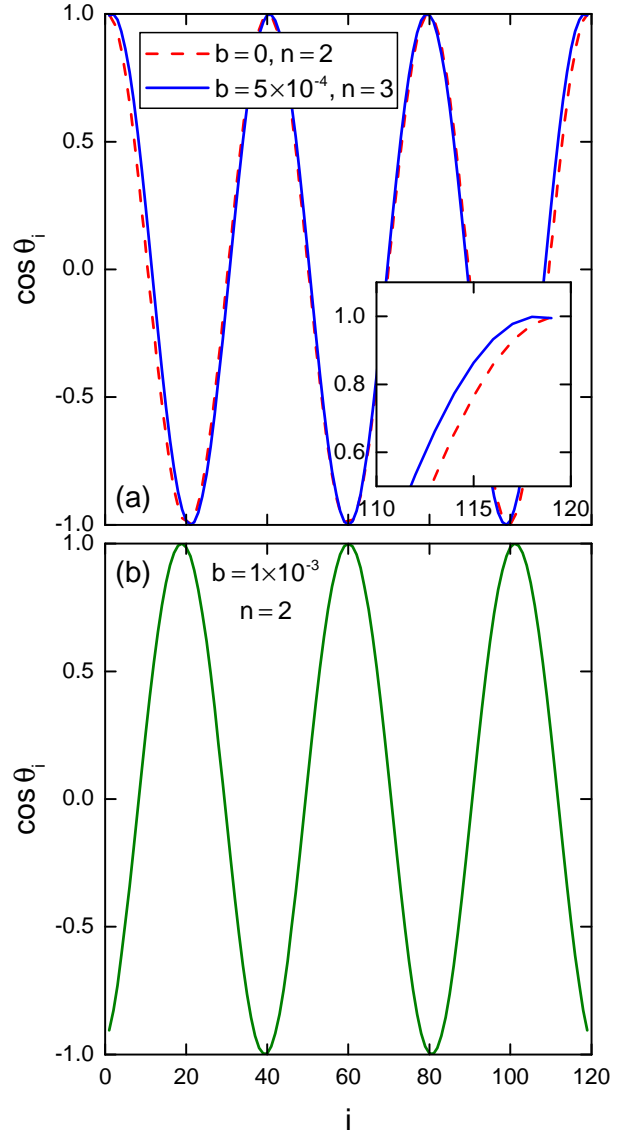}
\caption{(Color) Spin along the field direction as a function of site $i$ for $N=119$ and (a) $b=0$ and $0.0005$,
and (b) $b=0.001$.  Starting with $n=2$ complete periods at $b=0$, the spin configuration changes to $n=3$ at $b=0.0005$
and back to $n=2$ at $b=0.001$.  The increase in $n$ in (a) is produced by the downturn in $\cos \theta_i$ at the 
end sites, as shown in the insert.}
\end{figure}

Approximately 90\% of the old state is mixed with 10\% of the new state in order to avoid oscillations between possible solutions.
This procedure continues until no further updates are obtained.  In order to simplify this procedure, we use the fact that the spin state
is either mirror symmetric about the center for even $N$ or has spin angle $\theta_i = 0$ or $\pi $ at the central site for odd $N$.
These symmetry considerations reduce the number of spin degrees of freedom by half.
Even for the largest $N=250$ in Fig.1, our numerical procedure converges in less than $10^5$ time steps.

\begin{figure}
\includegraphics[width=8cm]{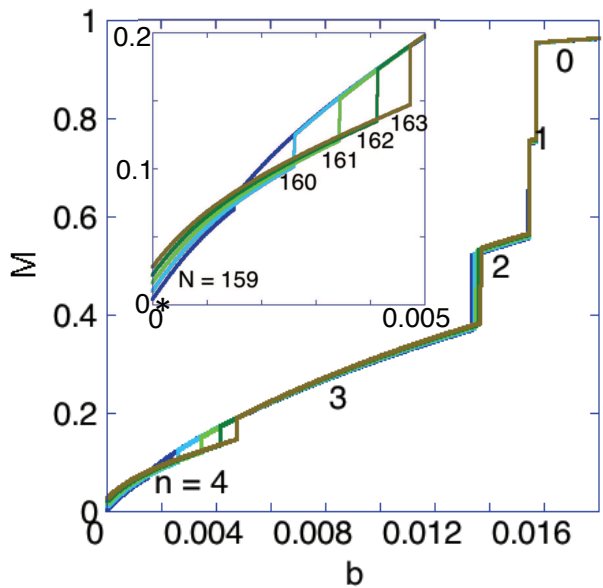}
\caption{(Color) Net average magnetization as a function of magnetic field for thicknesses $N=159$ to 163.  Inset shows the jump
from $n=2$ to $n=3$ for each thickness $N$.}
\end{figure}

We evaluate the phase diagram with $d=0.16$, which corresponds to a zero-field spiral with wavelength $\Lambda = 39.6$,
the same as observed in bulk Cr$_{1/3}$NbS$_2$ \cite{Togawa12}.
Phase boundaries are evaluated by either sweeping in thickness $N$ for a fixed magnetic field or in magnetic field $b$ for a fixed 
thickness.  As seen in Fig.2, these two techniques produce consistent results.  

In zero field, the spiral is not affected by the open boundary conditions because the rotation angle $\phi = \cos^{-1} (\vS_i \cdot \vS_{i+1}/S^2)$ 
does not change between pairs of spins at sites $i$ and $i+1$, even for the first and last pairs with $i=1$ and $N-1$.  Hence, the zero-field spiral has 
the bulk period $\Lambda =39.6$.  The transition from $n=n'$ to $n'+1$ complete periods occurs when $N-1$ just exceeds $(n'+1)\Lambda $.
So in zero field, the ferromagnetic state ($n'=0$) is stable below $N=41$.  

For any thickness, the spiral passes through states with $n=n_0$ to 0 complete periods in steps of 1 as $b$ increases.  
At high thicknesses ($N \gg \Lambda$), the ferromagnetic state is stable above $b_c \approx (\pi /4)^2d^2 = 0.0157$ \cite{Kishine05}.  
These results agree with earlier calculations \cite{Wilson13} made in the continuum limit and with previous results \cite{Kishine14} that assumed 
fixed boundary conditions.

The spin $\cos \theta_i$ along the field direction is plotted as a function of $i$ in Fig.3 for $b = 0.01$.
The pairs of thicknesses in Figs.3(a) and (b) were chosen so that the 
spiral passes from $n=n'$ to $n=n'+1$ with increasing $N$.   
In a nonzero magnetic field, the spin state becomes anharmonic with the spins spending more time
with $\cos \theta_i > 0$ than with $\cos \theta_i < 0$.

Surprisingly, each phase boundary in Fig.2 exhibits a slight bulge towards lower $N$ with increasing field.
Hence, the boundary between 0 and 1 node is 40.5 at zero field but drops to 39.5 at $b = 0.001$.   A value of 40.5
is recovered at $b = 0.003$.  The bulge moves to lower fields with increasing $N$.  Its presence implies that
a state with $n=n'$ complete periods at zero field can transform into a state with $n=n'+1$ before $n$
starts to decrease with increasing field.  To better understand this behavior, we plot the spin configuration for $N=119$ 
for fields $b$ ranging from 0 to 0.001 in Fig.4.  Although the number of complete periods $n$ increases from 2 to 
3 from $b=0$ to $0.0005$, the spin configuration is continuous.  On the other hand,
the spin configuration changes discontinuously (with $\cos \theta_i $ changing from $-1$ to 1 at the central site $i=60$)
as the field increases from $b=0.0005$ to $0.001$ and $n$ decreases from 3 back to $n=2$.  So the ``re-entrance" in 
Fig.2 is only a mathematical oddity and the lower transition has no physical effect.

Keeping the thickness fixed, we plot the average magnetization 
\begin{equation}
M=\frac{1}{N} \sum_{i=1}^N \cos\theta_i
\end{equation}
versus magnetic field $b$ in Fig.5.  As $b$ increases, the magnetization jumps 
when $n$ drops by one.  Those jumps become larger as $n$ approaches 0.  For thicknesses
$N=159$ thru 163, the spiral has 3 periods over the largest range of magnetic field.
Above $b_c$, the magnetization depends very weakly on field and continues to saturate
as $b$ increases. A close examination reveals that no jump occurs at the
``re-entrant" transition for $N=159$ at $b\approx 10^{-4}$, indicated by the star in the inset to
Fig.5.

Experimentally, a thin film will exhibit susceptibility peaks at the field separating $n'+1$ from $n'$ 
spiral periods.  For very thin films, there may be only one or two peaks.
For larger films, several peaks may be observable.  However,
the peaks will be largest as $n$ approaches 0. For 
films with $N-1$ below the spiral period $\Lambda $, the susceptibility will be a smooth function of field.

Because a realistic film will contain steps, 
its properties will be averaged over several thicknesses.  These steps will smear out the susceptibility peaks.  Nonetheless, 
Wilson {\it et al.} \cite{Wilson13} successfully observed two susceptibility peaks in a MnSi film.  Presumably, the lower peak marks the
transition from $n=2$ to 1 and the higher, larger peak from $n=1$ to 0. 

The insensitivity of the phase diagram of this model to the precise boundary conditions is remarkable.  Open
boundary conditions seem just as effective at quantizing the number of spiral periods as fixed boundary conditions.
Qualitatively, a phase diagram like that in Fig.2 requires only that the end spins are treated
differently than the interior spins.  Of course, the details of the phase diagram, such as the critical fields for the transitions 
from $n=n'$ to $n'-1$ for a fixed $N$, will depend on the specific boundary conditions.

To conclude, we have evaluated the magnetic state and phase diagram of a spiral with open boundary conditions.  For every thickness, the 
spin state passes through every number of complete periods from $n_0$ to 0 with increasing field.  A decrease in $n$ by one is marked by a jump
in the magnetization and by a peak in the magnetic susceptibility.  Our results indicate that the qualitative behavior of spiral states in thin films
does not depend on the precise boundary conditions.

We acknowledge helpful conversations with Zheng Gai.
Research sponsored by the U.S. Department of Energy, Office of Science, 
Office of Basic Energy Sciences, Materials Sciences and Engineering Division.

\end{document}